\documentclass[twocolumn,amsmath,amssymb,prd,nofootinbib]{revtex4}

\def\al{\alpha}
\def\be{\beta}
\def\ga{\gamma}
\def\de{\delta}
\def\ep{\epsilon}

\def\ze{\zeta}
\def\et{\eta}
\def\th{\theta}

\def\ka{\kappa}
\def\la{\lambda}

\def\rh{\rho}

\def\ta{\tau}

\def\ph{\phi}

\def\om{\omega}

\def\Th{\Theta}
\def\La{\Lambda}

\def\Ps{\Psi}
\def\Om{\Omega}
\def\cA{{\cal A}}
\def\cB{{\cal B}}
\def\cC{{\cal C}}

\def\cG{{\cal G}}
\def\cl{{\cal L}}
\def\cL{{\cal L}}

\def\fr#1#2{{{#1} \over {#2}}}

\def\frac#1#2{{\textstyle{{#1}\over {#2}}}}

\def\lsim{\mathrel{\rlap{\lower4pt\hbox{\hskip1pt$\sim$}}
    \raise1pt\hbox{$<$}}}
\def\gsim{\mathrel{\rlap{\lower4pt\hbox{\hskip1pt$\sim$}}
    \raise1pt\hbox{$>$}}}

\def\prt{\partial}

\def\etal{{\it et al.}}
\def\pt#1{\phantom{#1}}

\newcommand{\beq}{\begin{equation}}
\newcommand{\eeq}{\end{equation}}
\newcommand{\bea}{\begin{eqnarray}}
\newcommand{\eea}{\end{eqnarray}}
\newcommand{\bit}{\begin{itemize}}
\newcommand{\eit}{\end{itemize}}
\newcommand{\rf}[1]{(\ref{#1})}

\def\kr{{(k^{(4)})}{}}
\def\kdr{{(k^{(5)})}{}}
\def\kddr{{(k^{(6)}_1)}{}}
\def\krr{{(k^{(6)}_2)}{}}
\def\krrr{{(k^{(8)})}{}}

\def\kddrnp{{k^{(6)}_1}}%no parenthesis
\def\krrnp{{k^{(6)}_2}}
\def\krrrnp{  {k^{(8)}}  }

\def\kabc{ {\krrr_{\cA \cB \cC}} }

\def\kb{\overline{k}{}}
\def\kt{\widetilde{k}{}}

\def\kbrrr{ {(\kb^{(8)})}  }
\def\kbef{ {(\kb^{(8)}_{\rm eff})} }

\def\ktrrr{ {(\kt^{(8)})} }

\def\kbrrrnp{  {\kb^{(8)}}  }
\def\kbefnp{  \kb^{(8)}_{\rm eff} }

\def\kbabc{ {\kbrrr_{\cA \cB \cC}} }
\def\ktabc{ {\ktrrr_{\cA \cB \cC}} }

\def\mn{{\mu\nu}}
\def\ab{{\al\be}}
\def\abl{{\al\be\ldots}}
\def\abgd{{\al\be\ga\de}}

\def\klmn{{\ka\la\mu\nu}}
\def\ezet{{\ep\ze\et\th}}

\def\cG{{G}}

\def\Ra{R^\cA}
\def\Rb{R^\cB}
\def\Rc{R^\cC}

\def\sb{\overline{s}}

\def\cld#1{\cl^{(#1)}_{\rm LV}}

\def\mbf#1{\boldsymbol #1}

\begin{document}

\title{Anisotropic cubic curvature couplings}

\author{Quentin G.\ Bailey}

\affiliation{Physics Department, Embry-Riddle Aeronautical University,
Prescott, AZ 86301, U.S.A.}

\date{\today}

\begin{abstract}

To complement recent work on tests of spacetime symmetry in gravity, 
cubic curvature couplings are studied using an effective field theory description of spacetime-symmetry breaking.
The associated mass dimension $8$ coefficients for Lorentz violation studied do not result in any linearized gravity
modifications and instead are revealed in the first nonlinear terms in an expansion of spacetime around a flat background.
We consider effects on gravitational radiation through the energy loss of a binary system and 
we study two-body orbital perturbations using the post-Newtonian metric.
Some effects depend on the internal structure of the source and test bodies, 
thereby breaking the Weak Equivalence Principle for self-gravitating bodies.
These coefficients can be measured in solar-system tests, 
while binary-pulsar systems and short-range gravity tests are particularly sensitive.

\end{abstract}

\maketitle

\section{Introduction}
\label{intro}

General Relativity (GR) with the Einstein-Hilbert (EH) action can be expressed in the language of gauge field theory 
with the gauge symmetries of local Lorentz symmetry and diffeomorphism symmetry.
As such it is considered a non-abelian gauge theory, 
with nonlinear self-interaction terms in the interaction hamiltonian.
How well these interaction terms have been tested is in part measured by
what alternatives to the EH action are acceptable within experimental limits.
While using the EH action does not result in a renormalizable quantum field theory, 
the structure of the interaction terms can be tested in the classical limit through observation, 
gravitational waves, 
weak-field and strong-field tests of gravity.
These types of tests are particularly of interest precisely {\it because} the standard implementation of 
quantum field theory fails to yield stable radiative corrections for the EH action, 
or to put it simply, 
GR has not been combined with quantum theory in a satisfactory way.

One promising avenue for experimental probes is to test the very gauge symmetries, 
local Lorentz and diffeomorphism symmetry, 
upon which GR rests.
A strong possibility exists, 
in various theoretical scenarios in the literature, 
that these symmetries might not be exact in nature, 
particularly in the low-energy limit of an underlying unified theory \cite{strings, reviews}.
Due to this and other motivations, 
a broad program has been underway recently to identify
and measure possible spacetime-symmetry violations 
using a general effective field theory framework \cite{ck9798, k04}.
Though no significant positive signal yet exists, 
numerous areas involving gravity, 
electromagnetism, 
and other forces have placed strong limits \cite{tables}.
In the case of gravity, 
work is underway to identify and measure signals for spacetime-symmetry breaking
in the weak-field gravity regime.
Analysis has already been performed with lunar laser ranging \cite{battat07, bourgoin16}, 
atom interferometric gravimetry \cite{hm08ch09, mh1113},
gyroscopic tests \cite{beo13},
binary-pulsar tests \cite{shao14}, 
short-range gravity tests \cite{kl15, hust15, hust-iu16}
planetary ephemeris \cite{hees15}, 
cosmic rays \cite{kt15},
gravitational waves \cite{km16, yunes16}, 
and others \cite{gravexpt}.

The theoretical framework used for the analysis so far has been the linearized gravity limit 
of the Standard-Model Extension framework \cite{bk06, b0911, kt09, kt11, t12, bonder, bkx15, jty15}.
The flat spacetime version of the SME framework uses
a general lagrangian expansion with CPT and Lorentz-breaking terms formed from background tensor fields 
contracted with operators built from matter fields.
Since the introduction of nondynamical background fields in curved spacetime is generally incompatible with
Riemann geometry, 
an alternative approach is used where the Lorentz breaking is considered dynamical \cite{k04, bluhm15}.
This is the spontaneous Lorentz-symmetry breaking approach, 
which has been applied to the SME in the linearized gravity limit.

The linearized gravity limit, 
where the metric is expanded around a flat background,
suffices for many phenomenological applications since gravity 
is inherently weak in the solar system and it helps significantly to simplify calculations due to 
the complexity of handling spontaneous Lorentz-symmetry breaking in this context.
However, 
it is only in the quadratic and higher-order terms in the field equations of GR that we see the structure of the 
interaction terms dictated by the spacetime symmetries under study.
In light of this it is natural to extend existing SME analysis in this direction to test these interaction terms.

No analysis in the gravity sector of the SME framework beyond the linearized limit exists to date.
There are two aspects to this.
One, 
the construction of these higher-order terms is marred by complexity due to 
accounting for the dynamics of the fluctuations (Nambu-Goldstone and massive modes) 
and ultimately finding the correct energy-momentum tensor to higher order in metric fluctuations 
that is consistent with the underlying conservation laws, 
the known linearized results, 
and keeping generality \cite{b13}. 
Second, 
for most of the coefficients in the minimal sector of the (gravitational) SME, 
and many in the nonminimal sector, 
experiments can already be analyzed using linearized gravity results.
So in many cases it is not of phenomenological advantage to establish the nature of these terms.
However, 
it is of theoretical interest to establish the nature of such terms.
Furthermore, 
it may be the case that certain types of Lorentz violation can only be probed by looking 
beyond the linearized limit.

As a first step toward pushing the analysis beyond the linearized gravity regime, 
we examine terms in the nonminimal SME that provide an example of nonlinear, 
second order terms in the metric expansion around a flat background, 
and that are relatively straightforward to calculate.
Additionally, 
these terms are phenomenologically of interest since they do not appear
in linearized gravity and contain some interesting effects for experiments and observation as we show.
Note that alternatives to the approach discussed here include studying 
specific models of spontaneous Lorentz violation \cite{models, ms09}, 
phenomenological parametrizations of physics beyond GR \cite{parameter, myw12, yagi14}, 
or even alternative geometries \cite{finsler}.

The paper is organized as follows.
In Sec.\ \ref{theory}, 
we review the SME framework in the gravity sector and discuss 
the basic lagrangian and field equations for the cubic coupling to be studied in this work.
We study the possible effects of the modified field equations on aspects of 
gravitational radiation in Sec.\ \ref{radiation}.
The post-Newtonian metric and the associated two-body acceleration are the subject of 
Sec.\ \ref{PNlimit}.
In Sec.\ \ref{expt} we discuss the application of the two-body acceleration results 
to gravitational experiments and observation, 
including an estimate of the sensitivities to the relevant coefficients for Lorentz violation. 
The work is summarized in Sec.\ \ref{summary}.
Throughout this work we adopt the notational conventions of previous work in the gravity sector \cite{bk06, bkx15}.
In particular, 
we refer throughout the paper to the linearized, quadratic, 
and cubic limit in an expansion of the space-time metric $g_\mn$ around a flat background.
When needed this notion can be made precise by inserting $\ep<1$ in the expansion
for the metric $g_\mn=\et_\mn + \ep h_\mn$ so that linearized equations are $O(\ep)$, 
quadratic is $O(\ep^2)$, etc. 

\section{Theory}
\label{theory}

In the effective field theory description of local Lorentz violation in the gravity sector,
we write the Lagrange density of the underlying action as the sum of terms
\beq
\cl = \cl_{\rm EH} + \cl_{\rm LV} + \cl_k + \cl_{\rm M},
\eeq
where $\cl_{\rm EH}= \sqrt{-g}(R - 2 \La)/2\ka$
is the Einstein-Hilbert term with cosmological constant $\La$.
The term $\cl_{\rm LV}$ describes the (Lorentz-breaking) gravitational coupling to the coefficient fields,
$\cl_k$ contains the dynamics of the coefficient fields,
$\cl_{\rm M}$ describes the matter sector, 
and $\ka=8\pi G_N$.

The non-standard term $\cl_{\rm LV}$ can itself be written as a series 
involving covariant gravitational operators of increasing mass dimension $d$.
Each term is formed by contracting the coefficient fields $k_\abl$
with gravitational quantities 
including covariant derivatives $D_\al$ and curvature tensors $R_{\abgd}$.
The terms that have been studied previously include mass dimension $4$ through $6$ and are given by
the lagrange density 
\bea
\cl_{\rm LV} &=& \fr {\sqrt{-g}}{2\ka} [ \kr_\abgd R^\abgd + \kdr_{\al\be\ga\de\ka} D^\ka R^\abgd
\nonumber\\
&&
+\kddr_{\al\be\ga\de\ka\la} D^{(\ka} D^{\la)} R^\abgd
\nonumber\\
&&
+\krr_{\abgd\klmn} R^\abgd R^\klmn ],
\label{dim4-6lag}
\eea
where the parentheses indicate symmetrization with a factor of $1/2$.
The minimal SME is contained in the $d=4$ $\kr_\abgd$ case.
This term can be split into a total trace $u$,  
a trace $s_\mn$, 
and a traceless piece $t_\klmn$.
In the linearized gravity limit, 
these coefficients have already been explored theoretically and experimentally.
The mass dimension $5$ term $\kdr_{\abgd\ka}$, 
which breaks CPT symmetry, 
and the mass dimension $6$ terms $\kddrnp$ and $\krrnp$
has been studied in the context of short-range gravity, 
graviton vacuum \v{C}erenkov radiation, 
and gravitational waves \cite{bkx15, kt15, km16}.
It is clear that the terms with more than one power of curvature will contain higher than second
derivatives of the spacetime metric.
Lagrangian models with higher than second derivatives are known to have stability issues \cite{ost}.
However, 
in the SME we consider these terms as small corrections to the EH lagrangian, 
as part of a perturbative series around low energies, 
thus we consider the dynamics as primarily driven by second order differential equations \cite{km09}.

Our focus in this work is on the first cubic curvature term with mass dimension $d=8$, 
which lies beyond those considered so far in the SME expansion.
For simplicity from here onwards we neglect the other coefficients in the pure-gravity sector.
The $d=8$ term we shall consider is written as
\beq
\cld8 = \fr {\sqrt{-g}}{2\ka} \krrr_{\abgd\klmn\ezet} R^{\abgd} R^{\klmn}  R^{\ezet}.
\label{dim8lag}
\eeq
The coefficient fields $\krrr_{\abgd\klmn\ezet}$ have dimensions of length to the power $4$,
or inverse mass to the power $4$ in natural units.
Due to the contractions, 
the indices on the coefficients inherit the symmetries
of the Riemann tensor and it is totally symmetric in the 
groups of indices $\abgd$, $\klmn$, and $\ezet$.
Using the symmetry properties of the coefficients, 
we can determine that there are 1540 a priori independent components.
The coefficients $\krrr_{\abgd\klmn\ezet}$ naturally contain as a subset
various traced pieces that involve contractions of indices on the three Riemann tensors.
The cubic contractions of the Riemann tensor can be generally classified \cite{fulling92}, 
but we do not attempt to elucidate the traced pieces in Eq.\ \rf{dim8lag} in this work.

Toy models of spontaneous Lorentz violation that match the form of 
the lagrangian \rf{dim8lag} are straightforward to construct.
For example, 
for a model with a vector field $B_\mu$ one can consider a coupling of the form
$\cL \sim B_\mu B_\nu R^\mn B_\al B_\be R^{\al\ga} R^\be_{\pt{\be}\ga}$.
Such couplings will be constrained experimentally by analysis of the general $\krrrnp$ term.
We also note that there are indeed other mass dimension $8$ terms we shall not consider here. 
For example, one can form an operator with mass dimension $8$ by combining two covariant derivatives 
with two powers of the Riemann curvature tensor, $\cL \sim D_{(\mu} D_{\nu )} R_{\abgd} R_{\ezet}$.
However, 
such terms contribute to the linearized field equations, 
and have partly been studied previously in recent works \cite{kt15, km16}.

At this stage, 
we introduce a compact notation for the $\krrrnp$ coefficients and the curvature tensors.
We use the calligraphic capital letters $\cA$, $\cB$, etc.\ to denote the groups of Riemann-like indices:
$\cA=\abgd$, $\cB=\klmn$, and $\cC=\ezet$.
Thus for the curvature tensor we write $R^{\abgd}=\Ra$
and for the $\krrrnp$ coefficients we write $\krrr_{\abgd\klmn\ezet}=\kabc$.
Repeated indices indicate contraction across all four spacetime indices, 
so for example, $R_\cA \Ra =R_\abgd R^\abgd$.
When needed, some indices will be shown explicitly.

The first step for phenomenology is to extract the field equations
resulting from the lagrangian \rf{dim8lag}.
This can be achieved by variation with respect to the metric $g_\mn$ and the result is
\bea
G_\mn &=& \ka (T_M)_\mn + 6 D^{(\al} D^{\be )} \left( \krrr_{\al\mu\nu\be \cA \cB} \Ra \Rb \right)
\nonumber\\
&&
+ \frac 12 g_\mn \kabc \Ra \Rb \Rc 
\nonumber\\
&&
+ 9 R^{\al\be\ga}_{\pt{\al\be\ga}(\mu} \krrr_{\nu) \ga \al \be \cA \cB} \Ra \Rb
+ \ka (T_k)_\mn,
\label{fullfieldeqns}
\eea
where the energy-momentum tensor for matter is $(T_M)_\mn$ and
$(T_k)_\mn$ is the energy-momentum tensor contribution from the dynamics
of the coefficient fields $\kabc$ in the lagrangian term $\cL_k$.
We then assume assume an asymptotically flat background metric $\et_\ab$ as usual
and impose the conditions of spontaneous breaking of Lorentz symmetry.
In particular, it is assumed that the coefficient fields $\kabc$ have a vacuum expectation value
of $\kbabc$.
So for the metric and the coefficients fields we use 
the following expansions around the vacuum values:
\bea
g_\mn &=& \et_\mn + h_\mn,
\nonumber\\
\kabc &=& \kbabc + \ktabc.
\label{ssb}
\eea
Here $\ktabc$ represents the fluctuations around the vacuum value.\footnote{To keep track of the orders involved in the expansion
around a flat spacetime background, 
one can insert $\ep$ in front of $h_\mn$ and $\ktabc$ in Eq.\ \rf{ssb}}
In the chosen asymptotically flat cartesian coordinates, 
we assume the partial derivatives of the vacuum values vanish ($\prt_\mu \kbabc =0$).

Since the $\krrrnp$ lagrangian term is already of at least cubic order in the metric fluctuations $h_\mn$, 
the leading $\kabc$ terms in the field equations will be at quadratic order, 
as can be verified by insertion of the expansions \rf{ssb} into equation \rf{fullfieldeqns}.
If we confine attention to quadratic order field equations $O(h^2)$, 
the procedure for the elimination of coefficient fluctuations $\ktabc$ described in Refs.\ \cite{bk06, kt11, bkx15}
involves no fluctuations for $\kabc$ in the $\krrrnp$ terms in Eq.\ \rf{fullfieldeqns} 
because these contribute only at order $O(h^3)$.
We therefore obtain the modified Einstein equations valid to quadratic order as
\beq
G_\mn = \ka (T_M)_\mn 
+ 6 \kbrrr_{\al\mu\nu\be \cA \cB} \prt^\al \prt^\be \left( \Ra \Rb \right) + \ka (T_k)_\mn.
\label{fieldeqns}
\eeq
In Eq.\ \rf{fieldeqns},
we can take the curvature tensors in the second $\kbrrrnp$ term to be linearized in $h_\mn$, 
since the term is already at quadratic order.

As we are considering spontaneous symmetry breaking, 
the underlying action remains invariant under diffeomorphisms and local Lorentz transformations, 
and so the field equations will satisfy the conservation laws associated with these symmetries \cite{goldstone62}. 
In particular, 
the conservation laws associated diffeomorphism invariance holds, 
which can be checked as follows.
First note that the covariant divergence of the field equations \rf{fieldeqns} vanishes on the left side by the geometric identities
$D_\mu G^\mn=0$.
On the right-hand side, 
it can be verified that the divergence of the second $\kb$ term vanishes to quadratic order automatically.
This is in contrast to some of the other coefficients in the SME expansion \cite{bk06, bkx15},
where additional compensating terms (coming from an $\cL_k$-type term in the action) 
are necessary in the construction of the effective field equations. 
If we then assume that the matter tensor $(T_M)_\mn $ has independently vanishing divergence,
then it follows that the remaining piece $(T_k)_\mn$ must be independently covariantly conserved.
In addition to diffeomorphism symmetry, 
the conservation law for local Lorentz symmetry is also satisfied by virtue 
of the symmetry of the free indices of the terms on the right-hand side of \rf{fieldeqns}.

The second term in Eq.\ \rf{fieldeqns} is already in a form suitable 
for the calculation of the effects on the spacetime metric
since it only depends on the vacuum values of the coefficients $\kbrrrnp$, 
and the metric fluctuations $h_\mn$.
Therefore the final task in establishing the effective quadratic-order field equations
is to consider the unknown, 
independently conserved term $(T_k)_\mn$.
Considering the discussion above, 
this term is not necessary to ensure that the conservation laws hold at quadratic order.
Without knowing the explicit form for the terms in $\cL_k$, 
which would come from some underlying theory producing the coupling \rf{dim8lag}, 
we cannot directly calculate $(T_k)_\mn$.
One possibility is to study a large class of specific models of spontaneous Lorentz breaking
where the exact form of $\cL_k$ is known.
This task lies beyond the scope of the present paper but would be of interest for future study.
Note that investigations along these lines already exist for the coefficients $\kr_\abgd$.
For example, 
in a study of which vector models of spontaneous Lorentz breaking match the field equations presented in Ref.\ \cite{bk06},
it was shown that various assumptions, 
including assuming a vanishing independently conserved piece $(T_k)_\mn$,
restrict attention to a subset of all possible vector models \cite{ms09}.
We remark here that some preliminary analysis with vector models, 
with couplings of the form mentioned above, 
shows that the contribution of $(T_k)_\mn$ vanishes to quadratic order.

For the remainder of this work we shall adopt the assumption that $(T_k)_\mn=0$ to quadratic order, 
which is similar to an assumption adopted for extracting the field equations for the minimal SME coefficients $\kr_\abgd$.
Note that this assumption does not preclude the case when $(T_k)_\mn$ happens to be proportional to a term of the form
of the $\kbrrr$ term in equation \rf{fieldeqns} - in this case there is a rescaling of the field equations 
but the phenomenology remains the same.

\section{Radiation Effects}
\label{radiation}

If one considers the linearized limit of the field equations \rf{fieldeqns}, 
the $\kbrrrnp$ term on the right-hand side vanishes.
It is clear then that there are no effects on the propagator for the metric fluctuations $h_\mn$.
This implies, 
for example, 
that the propagation of gravitational waves is the same as in GR.
One consequence of this result is that the coefficients $\krrrnp$ lie beyond the reach of the previous analysis with quadratic actions 
\cite{km16} and analysis with modified dispersion relations \cite{myw12}.
We note, 
however, 
that we are assuming an expansion of spacetime around a flat background.  
If we were to generalize this to an expansion around a curved background, 
the possibility would then exist for linearized effects of the coefficients $\kbrrrnp$ to appear.
This could occur since the curvature terms in the lagrangian \rf{dim8lag} could take their background values plus fluctuations, 
leaving some terms quadratic in $h_\mn$ in the lagrangian.
This possibility has been considered by some authors studying Lorentz invariant cubic curvature couplings \cite{ccc}. 
In context of the lagrangian \rf{dim8lag}, 
this remains an open area of study beyond the scope of this work.  

Although there are no effects on the propagation of gravitational waves from the cubic curvature coupling terms $\kbrrrnp$, 
the appearance of these Lorentz-breaking terms at quadratic order in the field equations implies
the possibility of an effect on the energy-momentum loss of a system due to radiation from gravitational waves.
Indeed, 
the standard calculation in GR for the energy loss of a gravitational system involves using
the quadratic order terms in the usual Einstein equations \cite{mtw}.
To examine this possibility, 
we compute the total four-momentum and rate of four-momentum loss for a binary system using the field equations.
First we display the field equations in an alternative form:
\bea
(G_L)^\mn &=& \ka [(T_M)^\mn + \ta^\mn]
\nonumber\\
&&
+ 6 \kbrrr^{\al\mu\nu\be}_{\pt{\al\mu\nu\be}\cA \cB} \prt_\al \prt_\be \left( \Ra \Rb \right),
\label{relfe}
\eea
where $(G_L)^\mn$ is the linearized Einstein tensor, 
$\ta^\mn$ is defined by
\beq
\ta^\mn = \fr {1}{\ka} [(G_L)^\mn - G^\mn],
\label{pseudo}
\eeq
and the expression \rf{relfe} is valid to quadratic order in $h_\mn$.
Note also that to quadratic order, 
we raise and lower indices in the $\kbrrrnp$ term with the Minkowski metric $\et_\mn$ and its inverse.

The ordinary divergence $\prt_\mu$ of the right-hand side of \rf{relfe} vanishes and it can be interpreted
using an energy-momentum pseudotensor,
\beq
\Th^\mn=(T_M)^\mn + \ta^\mn+ \frac {6}{\ka} \kbrrr^{\al\mu\nu\be}_{\pt{\al\mu\nu\be}\cA \cB} 
\prt_\al \prt_\be \left( \Ra \Rb \right),
\label{em}
\eeq
for the total system of gravity plus matter.
One can define the total four-momentum $P^\mu$ of the system as the spatial integral of $\Th^{0\mu}$ over all space
in a fixed coordinate system.
For a generic spacetime, 
this integral does not converge.
However, 
we consider spacetimes that are asymptotically flat except for possible gravitational wave pieces 
in the metric fluctuations $h_\mn$.
Ordinarily in GR, 
the total four-momentum consists of two pieces: 
one due to the localized, 
or near zone solutions for the space-time metric, 
and the other due to possible gravitational radiation.
While the former can be shown to be constant in time, 
the latter can result in a time-varying total four-momentum 
on account of gravitational radiation carrying energy and momentum away from the system.

To see if there is a modification to the energy and momentum carried away from an isolated gravitational system,
we examine the rate of change of the spatial integral of $\Th^{0\mu}$ with respect to coordinate time $t$, 
or $dP^\mu /dt$.
After some manipulation using the conservation law $\prt_\mu \Th^\mn=0$, 
the expression can be turned into a surface integral over a sphere of radius $R$ centered on the gravitational system.
In this region, 
the matter energy momentum tensor vanishes and the only terms in $\Th^\mn$ that will survive in the limit $R \rightarrow \infty$
are quadratic in $h_\mn$, 
coming from wave solutions that fall off as $1/R$.

We shall restrict attention to results at leading order in the coefficients $\kbrrrnp$, 
which has the advantage that for the third Lorentz-violating term in Eq.\ \rf{em}, 
we can substitute GR results.
In particular, 
we use the result that the wave solutions for $h_\mn$ have a dependence on the retarded time $t_R=t-R$
with the wave vector $k_\mu = (1, n_j)$.
Here $n_j$ is an outward pointing normal from the origin
taken as the gravitational system's center of mass-energy. 
We then find that to leading order in the coefficients $\kbrrrnp$, 
\bea
\fr {dP^\mu}{dt} &=& -\int_S d^2 x \, n_j 
\Big[ \ta^{j\mu} 
\nonumber\\
&&
\hspace{-20 pt}
+\frac {24}{\ka} \kbrrr^{\pt{\al}j \mu}_{\al \pt{j\mu} \be l \ga \de m n \ep \ze p} k^\al k^\be k^\ga k^\de k^\ep k^\ze 
\fr {d^2}{d t_R^2} (\ddot{h}^{lm} \ddot{h}^{np}) \Big],\nonumber\\
\label{Pdot}
\eea
where a dot indicates a derivative with respect to $t_R$. 

We next examine the contributions of the two terms in \rf{Pdot} in turn.
Contributions to $h_\mn$ itself in the wave zone include the standard quadrupole terms in GR, 
higher-order terms in a post-Newtonian series, 
and possible contributions from the $\kbrrrnp$ terms in \rf{fieldeqns}.
These latter terms, 
however, 
can be expected to contribute only at second order in $\ep$ when considering 
an expansion around a flat background
$g_\mn=\et_\mn+\ep h_\mn$.
It would be of interest to calculate such terms to determine their possible effects 
on gravitational waves but this lies beyond the scope of the present work.
It suffices here to note that when inserted into \rf{Pdot} in the first term involving $\ta^{j\mu}$, 
these $O(\ep^2)$ corrections to $h_\mn$ will yield modifications only at third order in $\ep$ in the first term of \rf{Pdot}
since $\ta^{j\mu}$ is already at $O(\ep^2)$.
Thus if one neglects any contributions to $h_\mn$ in the wave zone from the $\kb^{(8)}$ coefficients,
the first term in \rf{Pdot} yields the standard result for energy and momentum loss from gravitational waves in GR. 
For example, 
for the energy loss one obtains the well-known quadrupole result 
$dP^0/dt = (G_N/5) \dddot{I}^{jk} \dddot{I}^{jk}$ where $I^{jk}$ is the (traceless) mass quadrupole tensor.

The second term in equation \rf{Pdot} is the leading Lorentz-violating correction to the energy-momentum loss 
for an isolated gravitational system.
At leading order in Lorentz violation we insert the standard quadrupole formula 
for the metric fluctuations $h^{jk}=2G_N \ddot{I}^{jk}/R$ into this term.
Focusing on a binary system, 
and using the leading expression for $I^{jk}$ for a slow-motion system, 
we obtain an explicit function of $t_R$ for this term that is periodic.
Furthermore, 
in this limit the $t_R$-derivative term does not depend on angles 
so the angular integral only applies to the projection of the $\kbrrrnp$ coefficients along $k^\mu$ 
and yields a linear combinations of these coefficients.
Upon time averaging the second term in \rf{Pdot} over one orbit, 
however, 
we obtain zero for this extra contribution to the energy-momentum flow.
We can trace this result to fact that a total $t_R$ derivative appears in the expression - 
when averaged over $t_R$ we end up evaluating a periodic function at the beginning and end of one cycle, 
thereby obtaining zero.
Therefore, 
to quadratic order in $h_\mn$, 
we can say that the $\kbrrrnp$ coefficients do not produce any leading-order effects for the energy-momentum loss
for gravitational waves.

\section{Post-Newtonian limit}
\label{PNlimit}

An alternative to exploring gravitational waves is to consider
the ``near zone" post-Newtonian effects on a gravitational system.
As usual, 
this involves an expansion in powers of the average speed $\bar v$ of the typical
body in the system with the Newtonian potential $U \sim v^2$ dominating over small relativistic corrections.
We employ a perfect fluid model to describe the bodies in the system, 
assuming the usual perfect fluid energy-momentum tensor for $(T_M)^\mn$.
Using this model, 
we solve the modified field equations to obtain the post-Newtonian metric
and ultimately the dynamical equations for a two-body system.

\subsection{Metric}
\label{metric}

In terms of mass density $\rh$, 
internal energy $\Pi$, 
pressure $p$, 
and four-velocity field $u^\mu$, 
the matter energy-momentum tensor is
\beq
(T_M)^\mn = (\rh+ \rh \Pi + p) u^\mu u^\nu + p g^\mn.
\label{fluidem}
\eeq
The Newtonian potential $U$, 
contained in $h_{00}/2$ satisfies Poisson's equation
\beq
-\vec \nabla^2 U = 4\pi G_N \rh.
\label{Newt}
\eeq
Assuming the matter is localized, 
the standard solution is
\beq
U (\mbf r, t) = G_N \int d^3r^\prime 
\fr{ \rh (\mbf r^\prime, t)}{ |\mbf r - \mbf r^\prime|}.
\label{U}
\eeq

The full post-Newtonian metric from \rf{fieldeqns} turns out to be that of General Relativity 
except for one additional term in the $O(v^4)$ piece of the metric components $h_{00}$, 
which we denote $\de h_{00}$.
The equation for this extra term can be obtained from the post-Newtonian expansion 
of the field equations \rf{fieldeqns}.
Focusing on solving for this piece, 
the relevant equation is that involving the $R_{00}$ component of the Ricci tensor, 
which is given to $O(v^4)$ by
\bea
R_{00} &=& \ka (S_M)_{00}+ 3 \kbrrr_{j00k \cA \cB } \prt_j \prt_k \left(\Ra \Rb \right)
\nonumber\\
&& +3 \kbrrr_{jllk \cA \cB }\prt_j \prt_k \left( \Ra \Rb \right),
\label{Ricci}
\eea
where the curvature components in the $\kbrrrnp$ term surviving at this post-Newtonian order 
include $R_{0j0k}$ and $R_{jklm}$.
The term $(S_M)_\mn$ is the trace-reversed energy-momentum tensor for matter.
The matter terms in $(S_M)_{00}$ along with other terms in $R_{00}$ contribute to the conventional GR
post-Newtonian metric.

To solve for the desired term $\de h_{00}$ we shall adopt a standard perturbative assumption 
and work to leading order in the coefficients for Lorentz violation.
We make the following coordinate choice on some of the components of the metric,
\bea
\prt_j g_{jk} &=& \frac 12 \prt_k (g_{jj}-g_{00}),
\nonumber\\
\prt_j g_{0j} &=& \frac 12 \prt_0 g_{jj},
\label{gauge}
\eea 
which matches earlier conventions \cite{bk06}.
It will be useful here to introduce a common shorthand for partial derivatives where $\prt_{jkl...}=\prt_j \prt_k \prt_l ...$.
The relevant equation becomes
\bea
-\fr 12 \vec \nabla^2 \de h_{00} =  48 (\kb^{(8)}_{\rm eff})_{jklmnp} \prt_{jk} (\prt_{lm} U \prt_{np} U).
\label{deheqn}
\eea
In this equation,
$\kbef_{jklmnp}$ are effective coefficients for Lorentz violation, 
given in terms of the underlying coefficients in the lagrangian by the expression
\bea
(\kb^{(8)}_{\rm eff})_{jklmnp} &=& \kbrrr_{0jk00lm00np0} + \kbrrr_{0jk0qlmq0np0} 
\nonumber\\
&&+\kbrrr_{0jk00np0qlmq}+\kbrrr_{qjkq0lm00np0} 
\nonumber\\
&&+\kbrrr_{0jk0qlmqrnpr} + \kbrrr_{qjkq0lm0rnpr}
\nonumber\\
&&+\kbrrr_{qjkqrlmr0np0}+\kbrrr_{qjkqrlmrsnps}.\nonumber\\   
\label{keff}
\eea
These effective coefficients $(\kb_{\rm eff})_{jklmnp}$ have symmetry 
in each of the pairs of indices $jk$, $lm$, and $np$.
Also, there is pairwise symmetry under the interchanges
$jk\leftrightarrow lm$, $jk \leftrightarrow np$, and
$lm \leftrightarrow np$, 
bringing the number of independent coefficient combinations in $\kbrrrnp$ to $56$.

We assume that the right-hand side of equation \rf{deheqn} represents a small correction to GR.
Using dimensional analysis this implies roughly that $\kb < L^4$, 
where $L$ is the typical length scale of the gravitational system, 
to be consistent with the perturbative assumption.
Proceeding, 
the Poisson-like equation \rf{deheqn} has the standard integral solution
\beq
\Ps = 48 (\kb^{(8)}_{\rm eff})_{jklmnp} \int d^3r^\prime 
\fr{ \prt^\prime_{jk} [\prt^\prime_{lm}U (\mbf r^\prime) \prt^\prime_{np} U (\mbf r^\prime)]}{4\pi |\mbf r - \mbf r^\prime|},
\label{Pot}
\eeq
where $\Ps=\de h_{00}/2$.
From dimensional analysis, 
the result of this integral could contain terms that vary with the inverse sixth power of the distance.
However, 
we shall show there are scenarios where the nonlinear nature of this potential 
yields terms that actually vary as the inverse cubic power of the distance, 
among other terms.

The integral in equation \rf{Pot} is taken over all space.
For a localized gravitational system, 
one can show convergence of the integral for large values of $r^\prime$ using the asymptotic
behavior of the Newtonian potential.
For small values of $r^\prime$, 
careful consideration is needed. 
The standard use of delta functions to describe the distribution of matter for point masses fails to give a convergent result
for \rf{Pot} and is therefore avoided in this treatment. 
Such subtleties arise even in the treatment of general relativistic terms at first post-Newtonian order, 
due to the nonlinear nature of gravity \cite{blanchet00, pw14}.
We will assume a sufficiently well-behaved mass density function $\rh (\mbf r^\prime)$.  
For example, 
to be integrable in the Newtonian potential $U$ (Eq.\ \rf{U}), 
$\rh$ must be at least a piecewise continuous function.

For calculations to follow, 
we develop the integral in \rf{Pot} further.
One convenient way to solve the integral in \rf{Pot} is first to express the Newtonian potentials 
in the integrand in terms of the mass density using \rf{U}.
Equation \rf{Pot} then involves three volume integrals.  
To proceed further we make use of a ``triangle function" which is a three-point function $\cG ({\mbf r},{\mbf y_1}, {\mbf y_2})$ 
defined by
\beq
\cG ({\mbf r},{\mbf y_1}, {\mbf y_2}) =  \fr {1}{4 \pi} \int d^3r^\prime 
\fr {1}{ |\mbf r - \mbf r^\prime| |\mbf r^\prime - \mbf y_1| |\mbf r^\prime - \mbf y_2|}.
\label{3pt}
\eeq
The solution to this integral with the appropriate boundary conditions is known and we use the result
\beq
\cG ({\mbf r},{\mbf y_1}, {\mbf y_2}) = 1- \ln (r_1+r_2+r_{12}),
\label{3ptsol}
\eeq
where $r_1=|\mbf r- \mbf y_1|$, $r_2=|\mbf r- \mbf y_2|$, and $r_{12}=|\mbf y_1- \mbf y_2|$ \cite{patiw02, pw14}.
Using this function, 
the integral \rf{Pot} can be written
\bea
\Ps &=& 48 (\kb^{(8)}_{\rm eff})_{jklmnp} \int d^3 y_1 \int d^3 y_2 \rh (\mbf y_1, t) \rh (\mbf y_2, t)
\nonumber\\
&&
\pt{48 (\kb^{(8)}_{\rm eff})_{jklmnp}} \times
\prt_{jk l_1 m_1 n_2 p_2} \cG ({\mbf r},{\mbf y_1}, {\mbf y_2}). 
\label{Pot2}
\eea

The calculation of the remaining GR terms in the post-Newtonian metric proceeds as usual
from the field equation \rf{fieldeqns} using the coordinate choice \rf{gauge}.
The complete metric to first post-Newtonian order includes terms up to $O(v^4)$ in $g_{00}$, 
$O(v^3)$ in $g_{0j}$ and $O(v^2)$ in $g_{jk}$.
It is given by
\bea
g_{00} &=& -1 +2U + 2\ph -2U^2 + 2 \Ps,
\nonumber\\
g_{0j} &=&- \frac 12 (7V^j + W^j),
\nonumber\\
g_{jk} &=& \de_{jk} (1+2U),
\label{PNmetric}
\eea
where $V^j$, 
$W^j$, 
and $\ph$ are given by
\bea
V^j &=& G_N \int d^3r^\prime \fr{ \rh^\prime }{ |\mbf r - \mbf r^\prime|},
\nonumber\\
W^j &=& G_N \int d^3r^\prime \fr{ \rh^\prime (r-r^\prime)^j v^{\prime k} (r-r^\prime)^k}{ |\mbf r - \mbf r^\prime|^3},
\nonumber\\
\ph &=&  G_N \int d^3r^\prime \fr{ [\rh^\prime ( 2 v^{\prime 2} + 2 U^\prime+ \Pi^\prime) + 3p^\prime]} { |\mbf r - \mbf r^\prime|}, 
%2ph= 4Ph1+4Ph2+2Ph3+6Ph4
%ph=2Ph1+2Ph2+Ph3+3Ph4
%Ph1~rh v^2, Ph2~rh U, Ph3 ~rh Pi, Ph4 ~p
\label{GRpots}
\eea
and a prime denotes a dependence on the integration variables $r^{\prime j}$.
We see that for this subset of the SME in the lagrangian \rf{dim8lag}, 
the only contribution to the post-Newtonian metric is to $g_{00}$ at $O(v^4)$ comprised of the $\Ps$ potential.
 
\subsection{Binary system dynamics}
\label{dynamics}
 
Our goal is to find the equations of motion for a two-body system comprised of gravitationally bound, 
or otherwise, 
distinct bodies.
To do this we shall employ a standard method of modeling the bodies using the perfect fluid description, 
and ultimately integrating the acceleration density over a given body to find the equation of motion for its,
suitably defined, 
center of mass.
We use a special fluid density $\rh^*=\rh \sqrt{-g} u^0$ that satisfies
the continuity equation $\prt_0 \rh^* + \prt_j (\rh^* v^j)=0$ and define the mass of a body $a$ and its center of mass as 
\bea
m_a &=& \int_a d^3 r \rh^* (t, \mbf r),
\nonumber\\
\mbf r_a &=& \fr {1}{m_a} \int_a d^3r \rh^* (t, \mbf r) \mbf r.
\label{com}
\eea
Subsequent time derivatives of $\mbf r_a$ yield the velocity $\mbf v_a$ and acceleration $\mbf a_a$ of the center of mass.

The starting point for the integration of the fluid equations over the body $a$ is the acceleration given by
\beq
\mbf a_a =  \fr {1}{m_a} \int_a d^3r \rh^* (t, \mbf r) \fr {d \mbf v}{dt}.
\label{comacc}
\eeq
Into this equation, we insert an expression for $\rh^* d \mbf v / dt$ from the perfect fluid equations of motion
$D_\mu (T_M)^\mn =0$.
In particular, 
the spatial components $\nu=j$ of these equations yields the acceleration density
\bea
\rh^* \fr {d \mbf v}{dt} = \rh^* \mbf \nabla U - \mbf \nabla p - \mbf A_{\rm GR} +\rh^* \mbf \nabla \Ps.  
\label{fluid}
\eea
On the right-hand side, 
the first two terms are the Newtonian contributions to the acceleration density, 
while the third term contains the contributions from the GR post-Newtonian terms.
The latter expression can be found in equation $44$ of Ref.\ \cite{bk06} or in standard references
\cite{mtw, tegp}.
Of primary interest is the last term involving the gradient of $\Ps$, 
which contains the contributions from the $\kbefnp$ coefficients to the fluid motion.

Equation \rf{fluid} is inserted into \rf{comacc} to find the acceleration of the body $a$.
The details of this calculation for the Newtonian and GR terms can be found elsewhere \cite{pw14}.
The correction to the acceleration of a body $a$ coming from the $\Ps$ term is given by the integral
\beq
\de \mbf a_a = \int_a d^3 r \rh \mbf \nabla \Ps,
\label{dea1}
\eeq 
where $\Ps$ is inserted from \rf{Pot2}.
To evaluate this we insert the mass density functions for each body $a$, $b$, $c$, etc. into the integrals, 
as $\rh=\rh_a + \rh_b+ \rh_c...$.
Note that these densities are localized in the neighborhood of each body and hence affect the 
domains for the volume integrals.
This procedure breaks up the three integrals over the variables $r$, $y_1$, $y_2$ 
into regions over different bodies ($abb$, $aab$, $abc$, etc.).
In particular, 
there will be an integral in which the three volume integrals are all over body $a$.
This we identify as the ``self-acceleration" term - it vanishes by the symmetries of the integral, 
which offers a consistency check with the energy-momentum conservation law discussion in Sec.\ \ref{theory}.
For the remaining terms, 
due to the seven partial derivatives appearing, 
a complete solution even for the case of widely separated bodies 
with negligible multipole moments, 
and neglecting tidal forces, 
is lengthy, 
although straightforward to compute using known methods.
We can classify the terms that appear for a two-body system by their dependence on the inverse power
of the relative distance between the two bodies $a$ and $b$, $\mbf r = \mbf r_a - \mbf r_b$, 
where $\mbf r_a$ and $\mbf r_b$ are the center of mass positions.

We focus on a two-body system and seek terms in \rf{dea1} with the least inverse powers of $r$. 
It turns out that a tractable expression for the two-body acceleration
with $r^{-4}$ and $r^{-6}$ dependencies appears and contains a novel dependence on the structure of the bodies. 
To obtain results along these lines, 
an efficient way to deal with the three-point function $\cG$  that is appropriate for widely separated bodies is needed.
For example, 
when the variables ${\mbf r}$ and ${\mbf y_1}$ lie in one body, 
and ${\mbf y_2}$ lies in a different body,
we can expand the logarithm in powers of $r_1/r_2$, 
to obtain
\bea
\cG ({\mbf r},{\mbf y_1}, {\mbf y_2}) &=& 1-\ln 2 -\ln r_2 - \fr {r_1}{2} \left( \fr {1}{r_2} + r_2 \hat{n}^j_1 \prt_j \fr{1}{r_2} \right)
\nonumber\\
&&
+\fr {r_1^2}{4} \left( \hat{n}^j_1 \prt_j \fr {1}{r_2} + \fr{r_2}{2} \hat{n}^j_1 \hat{n}^k_1 \prt_{jk} \fr{1}{r_2} \right)+...,\nonumber\\
\label{3ptexp}
\eea
where the unit vector $\mbf n_1$ points in the direction of $\mbf r_1$.
More details about this type of expansion can be found in Ref.\ \cite{patiw02},
where it was used for some terms in higher post-Newtonian GR.

For this calculation, 
we assume perfectly spherical bodies and ignore multipole moments and tidal terms over each body.
However, 
we do not ignore the finite size of each body, 
as this plays a critical role in the result.
After some calculation, 
the expression for the two-body acceleration of body $a$ is
\beq
\mbf a_a = - \fr {G_N m_b \mbf n}{r^2} + (\mbf a_{\rm GR})_a +\de \mbf a_a,
\label{totacc}
\eeq
where the first term is the Newtonian acceleration and $\mbf n=\mbf r /r$.
The second term contains the contributions from GR to $O(v^4)$:
\bea
(\mbf a_{\rm GR})_a &=& - \fr {G_N m_b}{r^2} \Big[ \mbf n\Big( v^2_a - 4 \mbf v_a \cdot \mbf v_b +2 v^2_b 
\nonumber\\
&&
-\frac 32 (\mbf n \cdot \mbf v_b)^2 
-5 \fr {G_N m_a}{r} - 4 \fr {G_N m_b}{r} \Big)
\nonumber\\
&&
-\mbf n \cdot (4 \mbf v_a - 3 \mbf v_b) (\mbf v_a - \mbf v_b) \Big]
\label{GRacc}
\eea
The Lorentz-breaking piece of the acceleration is given by
\bea
\de a_a^j &=& -576 \pi G_N^2 m_b (P_a + P_b) \kbef_{kl(mmnn)} \fr {n^{<jkl>} }{r^4}
\nonumber\\
&&
-15120 G^2_N m_b [ ({\tilde P}_a + {\tilde P}_b) \kbef_{klmnpp}
\nonumber\\
&&
+({\tilde P^\prime}_a + {\tilde P^\prime}_b) \kbef_{klmpnp}]
\fr { n^{<jklmn>} }{r^6},
\label{2bacc}
\eea
which is valid up to terms of order $1/r^7$.

The directional dependence in \rf{2bacc} is encoded in each of the totally symmetric 
and trace free combinations of unit vectors $n^{<jkl>}$ and $n^{<jklmn>}$ that depend 
on the unit vector $\mbf n$ and the kronecker delta $\de^{jk}$.
Such terms are readily constructed and formulas can be found in the literature \cite{ww96, pw14}.
Explicitly they are given by
\bea
n^{<jkl>} &=& n^j n^k n^l - \frac 15 (n^j \de^{kl} + n^l \de^{jk} +n^k \de^{lj}) 
\nonumber\\
n^{<jklmn>} &=& n^j n^k n^l n^m n^n - \frac 19 ( n^j n^k n^l \de^{mn} + {\rm perms} )
\nonumber\\
&&
+ \frac {1}{63} ( n^j \de^{kl} \de^{mn} + {\rm perms} ),
\label{stf}
\eea
where ``perms" indicates all independent permutations of indices;
$10$ total permutations for the second term in $n^{<jklmn>}$
and $15$ for the third term.
The internal terms in \rf{2bacc} are integrals for each (spherical) body given by
\bea
P_a &=& \fr {1}{m_a} \int_a d^3r \rh_a^2,
\nonumber\\
{\tilde P}_a &=& \fr {1}{35 m_a} 
\left( 8 \pi \int_a d^3r \rh^2_a r^2 + 46 \fr {\Om_a}{G_N} \right),
\nonumber\\
{\tilde P}^\prime_a &=& \fr {1}{35 m_a} 
\left( 16 \pi \int_a d^3r \rh^2_a r^2 -48 \fr {\Om_a}{G_N} \right),
\label{Pa}
\eea
where the same expressions hold for body b.
Here $\Om_a$ is the Newtonian self-energy of the body:
\beq
\Om_a = - \fr {G_N}{2} \int_a d^3r \int_a d^3r^\prime \fr {\rh \rh^\prime}{ |\mbf r - \mbf r^\prime|}.
\label{Om}
\eeq

Certain features of this acceleration are striking.
Firstly, 
there is a dependence on the structure of the two bodies via the integrals \rf{Pa}.
In particular, 
body $a$'s acceleration due to body $b$ depends on integrals of the density over each body.
This implies that the way in which the matter in each body is distributed affects the way 
it falls in the presence of another body, 
even in the limit of vanishing multipole moments and tidal forces,
thereby violating the Weak Equivalence Principle for gravitationally bound systems (GWEP) \cite{nord, tegp}.
Note that in deriving the two-body acceleration for the GR terms \rf{GRacc} it is necessary to 
impose internal equilibrium conditions on each body to eliminate the dependence of the acceleration 
on internal structure integrals over each body.
These virial conditions involve the pressure, 
internal velocity, 
and internal gravitational potential energy of each body.
Additionally, 
the mass of each body is renormalized to include the total internal energy.
Virial conditions were necessary, 
for example, 
to derive the many-body equations for the mass dimension $4$ coefficients $\sb_\mn$ to show 
that they satisfied GWEP to post-Newtonian order $O(v^3)$ \cite{bk06}.
In the present case of the $\kbefnp$ coefficients, 
we find that such virial conditions or mass renormalization cannot be used to eliminate or simplify the Lorentz-violating piece 
of the acceleration \rf{2bacc}, 
despite the appearance of internal structure integrals \rf{Pa}.

The particular nature of the GWEP violation is novel as well.
To illustrate this we will focus on just the dominant inverse quartic term in the modified acceleration \rf{2bacc}.
Suppose the two bodies have uniform densities and radii $a$ and $b$.
Let the masses of each body be $m_a$ and $m_b$ with total mass $M=m_a + m_b$. 
We define a weighted inverse radius for the bodies via
\beq
\fr {1}{{\bar R}^3} = \fr {1}{M} \left( \fr {m_a}{a^3} + \fr {m_b}{b^3} \right).
\label{meanR}
\eeq
With these definitions, 
the modification to the relative acceleration becomes
\beq
\de a^j \approx -432 \fr {(GM)^2}{{\bar R}^3} \fr {K_{kl} n^k n^l n^j - \frac 25 K_{jk} n^k}{r^4}.
\label{2bacc2}
\eeq
The coefficients $K_{jk}$ are the traceless combinations
\bea
K_{jk} &=& \fr 13 [2(\kb_{\rm eff})_{jklmlm}+(\kb_{\rm eff})_{jkllmm}] 
\nonumber\\
&& -\de_{jk} \fr 19 [2(\kb_{\rm eff})_{llmnmn}+(\kb_{\rm eff})_{llmmnn}].
\label{kappa}
\eea
We can now see directly the dependence on the size of the bodies in the system.
The more compact the bodies, 
the stronger the amplitude of the symmetry breaking signal.
Among the strongest sources will be binary pulsar systems and binary white dwarf systems.

The second aspect of the result, 
evident in either \rf{2bacc} or \rf{2bacc2}, 
is the anisotropy of the acceleration.
This is a ubiquitous feature of Lorentz-symmetry breaking.
It implies the acceleration generally points in a different direction from the line between the two bodies.
Note that the directional and inverse quartic behavior of the dominant Lorentz-breaking acceleration term 
does resemble an effective quadrupole contribution
($Q_{jk} \sim K_{jk}$).
However, 
$K_{jk}$ is taken as a fixed background field in the gravitational system while 
$Q_{jk}$ is dependent on the orientation of each body and the distribution of matter.
Note also that while the components of $Q_{jk}$ decrease when mass is concentrated toward the center of the body, 
the internal terms $P_a$ and $P_b$ increase.
Nonetheless, 
there can be a significant correlation with the quadrupole acceleration term from Newtonian physics
which should be taken into account for phenomenology.

As a caution, we note that we cannot apply the results \rf{2bacc} or \rf{2bacc2} to the case of black-hole orbits.
Since we are working within the post-Newtonian limit, and using the approximate quadratic-order field equations, 
the results do not apply for black hole solutions.
Furthermore, 
even if we attempted to approximate the full solution for the spacetime metric for large distances,  
we assumed for the derivation above that the mass density $\rh$ is bounded, 
which is inconsistent with black hole case. 
A separate investigation for the case of black holes,
with the complete field equations without any weak field assumptions,
remains an open problem.
Even for neutron stars, 
our results will only be approximate as we use the first post-Newtonian approximation in our fluid model.
In some vector models of spontaneous Lorentz violation, 
it has been shown that a relativistic computation of the structure of a neutron star can play a 
strong role in determining accurate limits on Lorentz violation parameters from binary pulsar systems \cite{yagi14}.
This issue remains an open question in the SME for future work.

\section{Observation and Experiment}
\label{expt}

The result for a two-body system in equation \rf{2bacc} can be used to calculate observable deviations from 
conventional orbits in GR.
If we focus on secular changes in orbital elements, 
which can be measured in binary pulsar system orbits and solar-system tests, 
we can calculate directly from equation \rf{2bacc} using standard methods.
For simplicity of presentation here, 
we omit the $O(1/r^6)$ and higher terms in the acceleration expression \rf{2bacc} and
use the truncated version \rf{2bacc2}. 
After some calculation we find no change in the semi-major axis $a$ and eccentricity $e$ 
of a Keplerian ellipse when averaged over one orbit.
The inclination $i$, 
angle of nodes $\Om$, 
and periastron angle $\om$ change according to 
\bea
\left< \fr {di}{dt} \right> &=& \fr {432 a n^3 (\cos \om K_{Pk} - \sin \om K_{Qk})}{5 (1-e^2)^2 {\bar R}^3},
\nonumber\\
\left< \fr {d\Om}{dt} \right> &=&  \fr {432 a n^3 \csc i (\sin \om K_{Pk} + \cos \om K_{Qk})}{5 (1-e^2)^2 {\bar R}^3},\nonumber\\
\left< \fr {d\om}{dt} \right> &=&  \fr {216 a n^3}{5 (1-e^2)^2 {\bar R}^3} [K_{PP}+K_{QQ}-2K_{kk} 
\nonumber\\
&&-2\cot i( \sin \om K_{Pk} + \cos \om K_{Qk}) ],
\label{prec}
\eea
where $\mbf P$, $\mbf Q$, and $\mbf k$ are three unit vectors describing the orientation of the orbit 
($\mbf P$ is along the peristron, $\mbf k$ is normal to the orbital plane, and $\mbf P \times \mbf Q= \mbf k$),
and $n$ is the angular frequency of the orbit.
In this expression we have projected the coefficients $K_{jk}$ along these unit vectors \cite{bk06}.
The last orbital element is the mean anomaly $l_0$.  
It is straightforward to calculate but omitted here since it typically does not impact phenomenology.

The results in \rf{prec} bear a close resemblance to the precessions in the minimal SME from
the coefficients $\sb_\mn$ and $a_\mu$ \cite{bk06, kt11}.
In fact, 
for small eccentricity, 
the combination of coefficients $\sin \om K_{Pk} + \cos \om K_{Qk}$ matches 
the combination of $\sb_\mn$ coefficients $\sin \om \sb_{Pk} + \cos \om \sb_{Qk}$
in the expressions for the secular change in the angles $\Om$ and $\om$ but with a different amplitude.
To estimate the sensitivity of orbital analysis to the five measurable coefficients $K_{jk}$, 
we can use the limits placed on the spatial $\sb_{jk}$ coefficients from planetary ephemeris
and binary pulsar analysis \cite{shao14, hees15}.
We use the order of magnitude expression $\sb_{jk} \sim K_{jk} \times 170 n^2 a /{\bar R}^3$ and assume
$\sb_{jk}$ is limited at the $10^{-10}$ level,
to obtain the crude estimates in the Table \ref{est}.
For solar system ephemeris tests, 
the factor in front of $K_{jk}$ differs by about 4 orders of magnitude between Mercury's orbit and Saturn's orbit, 
with Mercury having the largest factor due to its compactness.
We adopt the value for Earth, 
for which $\sb_{jk} \sim K_{jk} \times 10^{-31} \, {\rm km}^{-4}$.
For binary pulsar tests, 
the situation is much more favorable due to compactness of pulsars, 
and we find $\sb_{jk} \sim K_{jk} \times 10^{-15} \, {\rm km}^{-4}$ for typical binary pulsar systems.
Using multiple orbits or binary pulsar systems oriented differently, 
one can disentangle the different components of $K_{jk}$ 
(referring them to a standard coordinate system - the Sun-centered Celestial equatorial system), 
as was done for the minimal coefficients $\sb_\mn$ and $a_\mu$.
Additional components of $\kbef_{jklmnp}$ are also of interest, 
and the secular precessions can readily be calculated using \rf{2bacc},
with sensitivity suppressed by a factor of roughly $({\bar R}/r)^2$ relative to $K_{jk}$.

\begin{table}
\caption{\label{est} Estimated sensitivity level for different test scenarios.  In the case of solar system and binary pulsar tests, 
the sensitivity is to the $K_{jk}$ subset of the $(\kb^{(8)}_{\rm eff})_{jklmnp}$ coefficients.  For gravimeter and short-range tests, 
the sensitivity refers to the relevant components of $(\kb^{(8)}_{\rm eff})_{jklmnp}$.}
\begin{ruledtabular}
\begin{tabular}{cc}
Test & Sensitivity to $k^{(8)}$ (in ${\rm km}^4$)\\
\hline
solar system & $10^{21}$\\
gravimeter & $10^{13}$\\
binary pulsar & $10^{5}$\\
short range & $10^{-1}$\\
\end{tabular}
\end{ruledtabular}
\end{table}

Also appearing in Table \ref{est} are estimated sensitivities for short-range tests of gravity and 
Earth laboratory gravimeter tests.
Rather than the specific result in \rf{2bacc}, 
we use dimensional analysis based on the general integral expression for acceleration \rf{dea1}
to estimate these sensitivities.
This is because the terms in \rf{2bacc} varying with higher powers of the inverse distance between the
two masses will play a crucial role once the radius of the bodies and the interbody distances are comparable
(${\bar R} \sim r$).
The ratio of the modified acceleration $\de a$ to the Newtonian acceleration is approximately
$G_N m 48 \kbefnp /(c^2 L^5)$ where $L$ is the length scale of the experiment, 
and $c$ is the speed of light inserted for the proper units of $\kbefnp$.
Using this crude estimate, 
we find that short-range gravity tests
are likely to be the most sensitive to the coefficients in the modified acceleration \rf{dea1}.
Compared to orbital tests, 
the masses involved are miniscule in short-range gravity tests (about $1 {\rm g}$), 
but the distances are vastly smaller and the force can vary as the inverse seventh power of distance
by dimensional analysis - thus strengthening the amplitude of the signal.
For gravimeter tests, 
the source is the Earth and distances are much larger so we don't expect these tests to be as sensitive.
Note that for short-range gravity tests and free-fall experiments near the Earth, 
the full expression for the acceleration is needed and the approximate result \rf{2bacc} does not suffice.  

It is useful to consider how the signal from General Relativity compares in these scenarios.
Roughly speaking, 
the ratio of the first post-Newtonian acceleration in GR to the Newtonian acceleration 
is given by the dimensionless factor $G_N m/(c^2 r)$, 
as can be seem from \rf{GRacc}.
This factor is miniscule for laboratory masses ($\sim 10^{-28}$) and still small but in principle observable 
for gravimeter tests ($\sim 10^{-10}$) \cite{bct77}.
However, 
the modified (point-mass) acceleration in GR is either proportional to velocity terms which are negligible, 
or scales the usual Newtonian acceleration and is therefore unlikely to interfere 
with Earth-laboratory tests seeking the Lorentz-breaking acceleration in \rf{dea1}.

For orbital tests, 
analysis can proceed using \rf{2bacc} or \rf{2bacc2} so long as
the interbody distance is sufficiently large compared to the size of the bodies in the system.
When considering short-range gravity tests \cite{SR}, 
it is necessary to work out the full integral in \rf{dea1}, 
along with the Newtonian force to obtain the needed total force between two laboratory test bodies.
However, 
analysis usually proceeds by using the point-mass force expression in a numerical integration code, 
along with modeling of the experiment.
In the present case, 
the point mass expression cannot be defined properly.
The starting point instead would be the integral expression for the acceleration of body $a$ 
due to body $b$, 
displayed in equation \rf{dea1}, 
which can be evaluated numerically for a given distribution of mass.
The seven partial derivatives inside the integral can be calculated in a straightforward manner
but result in a lengthy expression inside the integrand - this could also be implementation numerically.

In principle then, 
short-range gravity tests can measure a subset of the coefficients $\kbef_{jklmnp}$
in the laboratory frame.
As with other SME coefficients, 
they are considered constant in the canonical,
approximately inertial, 
Sun-centered frame (SCF) \cite{tables, sunframe}.
Neglecting boost effects from the Earth's velocity, 
a rotation $R^{jJ}$ dependent on the the Earth's sidereal frequency $\om_\oplus$
is needed to relate the lab frame to the SCF.
Specifically, 
the coefficients are transformed according to 
\beq
\kbef_{jklmnp} = R^{jJ} R^{kK} R^{lL} R^{mM} R^{nN} R^{pP} \kbef_{JKLMNP},
\label{rot}
\eeq
where the components in the SCF are denoted with capital letters.
The laboratory-frame coefficients are thus time dependent
and can potentially modulate the measured force with up to $6$th order harmonics
in the frequency $\om_\oplus$.
As with the mass dimension $6$ coefficients explored in Ref.\ \cite{bkx15}, 
the relevant short-range tests are those that satisfy the perturbative criteria, 
which implies that sensitivity at the level of the Newtonian force between 
the masses is needed.

What can be said about the possible sizes of the coefficients $(\kb^{(8)})_{(a)(b)(c)}$?
The SME effective field theory framework describes a broad class of possible effects, 
and does not make specific predictions concerning the sizes of these coefficients.
However, 
the inherent weakness of gravity compared to the other forces in nature evidently leaves room
for violations of space-time symmetry that are large compared to other sectors.
Consider the current limits on coefficients in the gravity sector.
For the coefficients $\sb_\mn$, 
the best laboratory limits are at the $10^{-10}$ level, 
with improvements of up to four orders of magnitude in astrophysical tests on these dimensionless
coefficients \cite{kt15}.
However, 
for the mass dimension $6$ coefficients $(\kb^{(6)}_1)_{\ka\la\mu\nu\al\be}$ and $(\kb^{(6)}_2)_{\ka\la\mu\nu\al\be\ga\de}$, 
the limits are at the $10^{-8} \, {\rm m}^2$ level.
When compared to the Planck length $10^{-35} \, {\rm m}$, 
it is evident that symmetry breaking effects that are not Planck suppressed could still have escaped detection.  
This kind of ``countershading" occurs for matter-gravity couplings such as the $a_\mu$ coefficients 
and in other sectors \cite{kt09, counter}.  
This theme continues for the mass dimension $8$ coefficients $\kbrrrnp$ in the cubic couplings considered in this work, 
where coefficients as large as $1 \, {\rm km}^4$ could have escaped detection.
Note also that any particular model that matches the form of the lagrangian \rf{dim8lag} will be subject to any limits 
garnered from the analysis herein.  

\section{Summary}
\label{summary}

In this work we studied a cubic curvature coupling describing general Lorentz and diffeomorphism symmetry breaking for gravity, 
as part of the general effective field theory expansion of the SME.
The basic lagrangian for this coupling is given in equation \rf{dim8lag} and the degree of symmetry breaking is described 
by the set of coefficient fields $\krrr_{\abgd\klmn\ezet}$ with inverse quartic mass dimension.
We studied the field equations for the spacetime metric up to quadratic order in an expansion 
around a flat background \rf{fieldeqns} and assuming spontaneous Lorentz-symmetry breaking.
The cubic coupling term provides a readily calculable example of the effects of Lorentz and diffeomorphism symmetry breaking 
at second order in the metric fluctuations $h_\mn$ in the SME framework.
The field equations also satisfy the conservation laws expected of spontaneous symmetry breaking to quadratic order 
in the metric fluctuations $h_\mn$.

The remainder of the paper explored the phenomenological consequences of the field equations 
for gravitational waves and post-Newtonian physics.
In Sec.\ \ref{radiation}, 
we showed some key null results.
Firstly, 
the propagation and dispersion of gravitational waves is unaffected by the cubic coupling form 
of Lorentz breaking that we consider.
Second, 
the energy and momentum loss for a binary system radiating gravitational waves to spatial infinity 
remains standard upon averaging over an orbital time scale.

In Sec.\ \ref{PNlimit}, 
we focused on the weak-field, slow motion limit and derived the post-Newtonian metric in Eq.\ \rf{PNmetric}.
The cubic curvature coupling term in the SME results in an $O(v^4)$ correction to the metric components $g_{00}$ comprised
of the potential $\Ps$ given in Eq.\ \rf{Pot}.
We modeled matter as a perfect fluid and used this to derive the acceleration of a massive self-gravitating body in a two-body system.
The modification to the GR acceleration is given in Eq.\ \rf{2bacc} and contains GWEP-violating dependence on internal structure.
The Lorentz-breaking effects in the post-Newtonian limit are controlled by a subset $\kbef_{jklmnp}$ of $56$ combinations 
of the full $\krrr_{\abgd\klmn\ezet}$ set of coefficients from the lagrangian, 
thus making analysis in this limit more tractable.

In Sec.\ \ref{expt}, 
we focused on the two-body acceleration and considered observations with binary pulsars and solar system tests.
We found the strongest sensitivity is likely to be with short-range gravity tests involving controlled laboratory masses, 
since the strength of the signal grows dramatically with the decreasing separation of masses.
A crude estimate of the sensitivity for different categories of tests is provided in Table \ref{est}.
Similar to other SME coefficients, 
the cubic coupling coefficients exhibit a kind of countershading, 
where the weakness of gravity can hide comparatively large Lorentz violation.

\begin{acknowledgements}

I thank Brett Altschul and Michael Seifert for valuable comments on this manuscript.
This work was supported in part by the National Science Foundation
under grant number PHY-1402890.

\end{acknowledgements}

\end{document}